\documentclass[a4paper,aps,pre,superscriptaddress,floatfix,nofootinbib,longbibliography,notitlepage,onecolumn]{revtex4-1}
\usepackage{bbm}
\usepackage{bbold}
\usepackage{bbm}
\usepackage[pdftex]{graphicx}
\usepackage{latexsym,amsmath,verbatim,amssymb,txfonts}
\usepackage{lipsum}
\usepackage{color}
\usepackage{rotating}
\usepackage{verbatim}
\usepackage{multirow}
\usepackage[english]{babel}
\usepackage{comment}
\usepackage{bm}
\usepackage{amsmath}
\usepackage{amsfonts}
\usepackage{amssymb}
\usepackage{color}
\usepackage{braket}
\usepackage{blkarray, bigstrut}

\begin{document}

\title{Synchronization of Dirac-Bianconi driven oscillators}

\author{Riccardo Muolo}
\affiliation{RIKEN Center for Interdisciplinary Theoretical and Mathematical Sciences (iTHEMS), Saitama 351-0198, Japan}
\affiliation{Department of Systems and Control Engineering, Institute of Science Tokyo (former Tokyo Tech), Tokyo 152-8552, Japan}
\email{riccardo.muolo@riken.jp}

\author{Iván León}
\affiliation{Department of Applied Mathematics and Computer Science, Universidad de Cantabria, Santander 39005, Spain}

\author{Yuzuru Kato}
\affiliation{Department of Complex and Intelligent Systems, Future University Hakodate, Hokkaido 041-8655, Japan}

\author{Hiroya Nakao}
\affiliation{Department of Systems and Control Engineering, Institute of Science Tokyo (former Tokyo Tech), Tokyo 152-8552, Japan}
\affiliation{International Research Frontiers Initiative, Institute of Science Tokyo (former Tokyo Tech), Kanagawa 226-8501, Japan}

\begin{abstract}

In dynamical systems on networks, one assigns the dynamics to nodes, which are then coupled via links. This approach does not account for group interactions and dynamics on links and other higher dimensional structures. Higher-order network theory addresses this by considering variables defined on nodes, links, triangles, and higher-order simplices, called topological signals (or cochains). Moreover, topological signals of different dimensions can interact through the Dirac-Bianconi operator, which allows coupling between topological signals defined, for example, on nodes and links. Such interactions can induce various dynamical behaviors, for example, periodic oscillations. The oscillating system consists of topological signals on nodes and links whose dynamics are driven by the Dirac-Bianconi coupling, hence, which we call it Dirac-Bianconi driven oscillator. Using the phase reduction method, we obtain a phase description of this system and apply it to the study of synchronization between two such oscillators. This approach offers a way to analyze oscillatory behaviors in higher-order networks beyond the node-based paradigm, while providing a ductile modeling tool for node- and edge-signals.
\end{abstract}

\maketitle

\section{Introduction}

In complex systems, the dynamics is shaped by the interactions among the elementary units and such interactions have been mostly assumed to be pairwise (two-body). However, in recent years, increasing evidence emerged towards the fundamental role that higher-order interactions play in describing real-world systems~\cite{battiston2020networks,bianconi2021higher,bick2023higher,boccaletti2023structure}. This triggered the interest of scholars in nonlinear dynamics, particularly due to the interplay between dynamics and higher-order topology~\cite{natphys,muolo2024turing,millan2025topology}. In fact, it has been shown that higher-order interactions enrich the dynamics of the systems, with applications regarding synchronization~\cite{tanaka2011multistable,skardal2020higher,millan2020explosive,lucas2020multiorder,wang2021collective,skardal2021higher,gambuzza2021stability,gallo2022synchronization}, random walks~\cite{schaub2020random,carletti2020random}, pattern formation~\cite{carletti2020dynamical,muologallo}, chimera states \cite{kundu2022higher,muolo2024phase}, opinion dynamics~\cite{iacopini2019simplicial,deville2020consensus,ziegler2022balanced,schawe2022higher}, and control~\cite{de2022pinning,della2023emergence,muolo2025pinning}, to name a few. In all of the above works, the authors showed that certain dynamics are only made possible by the higher-order interaction topology, namely, hypergraphs and simplicial complexes, which are extensions of networks. This is particularly interesting, as the complexity of the system lies not in the model, but rather in the interactions between the elementary units, and even very simple settings yield rich dynamics when higher-order interactions are present \cite{kuehn2021universal,leon2024}.  

In the study of dynamics on networks, the classical approach relies on node-based dynamics, where to each node is associated a variable, e.g., the phase of an oscillator, or the state of a neuron. Higher-order interactions in the sense of many-body (group) interactions persist in such node-based paradigm: in fact, the dynamical variables remain anchored to the nodes, while what change is the interactions between them, mediated by hyperedges of higher orders, i.e., group interactions. However, there are systems which do not fit well in this paradigm, and some examples come from collaboration networks \cite{patania2017shape}, brain networks \cite{faskowitz2022edges}, climate science \cite{singh2020fingerprint}, and transportation networks \cite{katifori2010damage}. To describe and analyze such systems, scholars have developed a framework that goes beyond node-based dynamics, using tools from algebraic topology and exterior calculus \cite{bianconi2021higher}, which allow us to consider the dynamics of topological signals, i.e., state variables defined not only on the nodes, but also on links, triangles, and higher-order simplices. The structure on which the dynamics is defined is called simplicial complex, which is a particular case of hypergraph with a closure relation, i.e., each $(d+1)$-dimensional simplices needs to be bounded by all $d$-simplices, and so on. As a simple example, a $2$-simplex (i.e., a triangle), needs to be bounded by all $1$-simplices (i.e., the links), which have to be bounded by all the $0$-simplices (i.e., the nodes). In this notation, a network can be considered as a $1$-simplicial complex, i.e., a simplicial complex made only of $1$- and $0$-simplices. Moreover, though the Dirac-Bianconi operator \cite{bianconi2021topological}, variables on different simplices of $1$ dimension up or down can be coupled. In this new framework, several classical nonlinear dynamics on networks model, such as the Kuramoto model \cite{kuramoto1975}, Benjamin-Feir instability \cite{nakao2014complex}, the Master Stability Function \cite{fujisaka1983stability,pecora1998master}, and Turing pattern formation \cite{NM2010,muolo2024turing}, have been extended for topological signals on simplicial complexes, namely, the Topological Kuramoto model~ \cite{millan2020explosive,Calmon2021topological,ghorbanchian2021higher,li2023synchronization,arnaudon2022connecting,nurisso2024unified,wang2025higher}, Global Topological Synchronization \cite{carletti2023global,wang2024global,carletti2025global}, and Topological Turing patterns \cite{Giambagli2022diffusion,muolo2024three}. The study of patterns and oscillations of topological signals in higher-order simplices is particularly interesting for the study of edge-signals in the brain \cite{faskowitz2022edges}.

In this work, we introduce a setting that is complementary to the above works. Before that, let us take a step back to the node-based framework. When studying synchronization dynamics on networks, it is common to view each node as a self-sustained oscillator, i.e., with a stable limit cycle, and then consider the synchronization between them \cite{pikovsky2001synchronization}. When considering such systems, there is a powerful technique, called phase reduction, which allows {us} to reduce each high dimensional oscillator to a phase variable \cite{nakao16,pietras2019network}. In general, the application of phase reduction yields simple phase models that are easier to understand and that have been widely studied. In fact, the Kuramoto model is derived through phase reduction~\cite{Kuramoto_book} and provides a simple and general mechanism for the emergence of synchronization~\cite{Kuramoto_book,acebron2005kuramoto,rodrigues2016kuramoto}. In recent years, extensions of this method have been considered, such as corrections via an expansion of the (small) coupling strength~\cite{leon19,leon22a,Mau23,bick2023higher}, the dynamics of the amplitude variables under non-weak coupling~\cite{kurebayashi2013phase,Wilson_2016,shirasaka2017phase}, or mean field variables \cite{leon20}, which allowed {us} to overcome some of the limitations and extend its applicability. All the above references consider the nodes as oscillators. However, periodic dynamics can emerge also in more complex settings. For example, most neurons are not oscillatory, but oscillatory patterns are observed at larger scale \cite{wilson1972excitatory,brunel1999fast}. This can be modeled by a network of interacting excitable units, whose isolated dynamics are not oscillatory, but exhibit self-sustained oscillations as a result of the coupling \cite{kawamura2011collective,pazo2016quasiperiodic,devalle2017firing,nakao2018phase,politi2018collective,di2018transition,nakao2021sparse}. In particular, authors of Ref. \cite{nakao2018phase} derived a phase description of oscillatory networks and studied their synchronization through phase reduction theory. 

Inspired by the latter work, we further extend this framework by considering networks where the dynamical variables lie on the nodes and the links, and are solely coupled via the Dirac-Bianconi operator.  When such {systems} exhibit self-sustained oscillations, we call them \textit{Dirac-Bianconi driven oscillators}, because their periodic behavior is induced by the Dirac-Bianconi operator. We then give a full description of such structures and study their synchronization through phase reduction theory.

The paper is organized as follows. In the next Section, we will give the theoretical background on topological signals and the Dirac-Bianconi operator. In Section \ref{sec:dirac_osc}, we first introduce and characterize a Dirac-Bianconi driven oscillator, and then study numerically the synchronization of two coupled oscillators. Finally, in Sec. \ref{sec:phase_red_dirac}, we give a phase description of such systems that explains the numerical results.

\section{Theory of topological signals and the Dirac-Bianconi operator}\label{sec:dirac}

In this Section we present a basic introduction to simplicial complexes and topological signals, limiting ourselves to define what is needed to characterize Dirac-Bianconi driven oscillators. For a more detailed discussion that includes other important concepts, the interested reader may consult more specialized references \cite{bianconi2021higher,millan2025topology,Lim2020,schaub2021signal}.

A $d$-dimensional simplex is a set of $d+1$ nodes that models a higher-order interaction among $(d+1)$ entities or represents a discrete object of dimension $n$. A $0$-simplex is a node, a $1$-simplex is a link, a $2$-simplex is a triangle, and so on. The faces of a $d$-simplex are all its $(d-1)$-dimensional subsimplices. For instance, the faces of a triangle (2-simplex) are its three links (1-simplices), and each link has two node (0-simplex) faces. Except for the $0$-simplices, each simplex has an arbitrary orientation, i.e., each simplex is labeled by an ordered list of nodes. For example, the (symmetric) link connecting nodes $i$ and $j$ can be labeled as $[i,j]$; then $[j,i]=[i,j]$. Linear combinations of $d$-simplices are also called $d$-chains. A simplicial complex is built by attaching simplices by their faces. Note that, in a simplicial complex, the existence of a $d$-simplex implies the existence of all its faces. Given $N_k$ $k$-simplices and $N_{k-1}$ $(k-1)$-simplices, the boundary operator $\mathbf{B}_k$ is a $N_{k-1} \times N_k$ matrix defined by:
\begin{equation} \label{matrixBoundary}
    \mathbf{B}_k(i,j)=
    \begin{cases}
        1 & \text{if the $i$-th $(k-1)$-simplex is a face of the $j$-th $k$-simplex}\\
        & ~~~~~~~~~~~~~~~~~~~~~~~~~~~~~~~~~~~~~~~~~~~~~~~~~~~~~~ \text{with same orientation,}\\
        -1 & \text{if the $i$-th $(k-1)$-simplex is a face of the $j$-th $k$-simplex}\\
        & ~~~~~~~~~~~~~~~~~~~~~~~~~~~~~~~~~~~~~~~~~~~~~~~~~~~~~~ \text{with opposite orientation,}\\
        0 & \text{if the $i$-th $(k-1)$-simplex is NOT a face of the $j$-th $k$-simplex.}
    \end{cases}
\end{equation}

Note that the boundary operator $\mathbf{B}_k$ changes when the orientation is changed, but the structure of the simplicial complex does not change; one can understand this by thinking of Cartesian coordinates: the axis can be changed, but this does not change the structure of the space. Note also that, in the case of network, i.e., consisting only of $0$-simplices and $1$-simplices, $\mathbf{B}_1$ is the well-known incidence matrix \cite{newmanbook}.

We now define topological signals (also called cochains) as functions defined on simplices (chains). A $k$-dimensional signal is a function assigning a real value to each $k$-simplex $\sigma^i_k$. The values can be arranged in a vector\footnote{Note that the dimension of the topological signal and the dimension of the simplex are not related. As one can define an $n$-dimensional vector on the nodes ($0$-simplices), the same can be done on any $k$-simplex.} $\vec{x} \in \mathbb{R}^{N_k}$, where each entry is $x_i = x(\sigma^i_k)$. These signals must be orientation-sensitive, i.e.,  $x(\sigma_k) = -x(-\sigma_k)$, to ensure the invariance under the change of the arbitrary orientation \cite{carletti2023global}. {Note that $d$-cochains are the discrete equivalent of differential $d$-forms~\cite{Lim2020,grady2010discrete}. Hence, $0$-cochains naturally model scalar fields, $1$-cochains vector fields, e.g., the gradient of a scalar, $2$-cochains bivector fields, e.g., the curl of a vector~\cite{dorst2007geometric}.}

The dynamics of such signals is then given by:
\begin{equation}
    \dot{x}_i = f_i(x_i),
\end{equation}
where $f_i$ must be odd, i.e., $f_i(-x_i) = -f_i(x_i)$, to preserve the orientation invariance. 

The boundary operator \eqref{matrixBoundary} connects topological signals (cochains) of consecutive dimensions: $\mathbf{B}_k$ on a $k$-cochain $\vec{x}$, it gives a $(k-1)$-cochain $\vec{z} = \mathbf{B}_k \vec{x}$, while acting with the transpose gives a $(k+1)$-cochain $\vec{g} = \mathbf{B}_{k+1}^\top \vec{x}$. The transpose of the boundary is known as the coboundary operator. One cannot connect topological signals of dimensions $k$ and $k\pm 2$ because the (co)boundary of a (co)boundary is zero~\cite{bianconi2021higher}.

For sake of simplicity, let us consider the case of a $1-$simplicial complex, i.e., a network, hence, only $0-$simplices (nodes) and $1-$simplices (links) are involved, with their respective topological signals\footnote{Note that the terms \textit{cochains}, which we will mainly use from now on, and \textit{topological signals} are equivalent.}, namely, $0-$cochains $\vec{u}$ (state variables on the nodes) and $1-$cochains $\vec{v}$ (state variables on the links). Note that the same formalism can be used to consider any dimension, as long as they are consecutive, given that Dirac-Bianconi operator connects topological signals of consecutive dimension. The state of the system is described by the topological spinor $\vec{w}$, defined as \begin{equation}
    \vec{w}=\begin{bmatrix}
        \vec{u} \\ \vec{v}
    \end{bmatrix}.
\end{equation} The topological spinor $\vec{w}$ has a $0-$cochain as a first component and a $1-$cochain as a second component. 

To couple cochains of adjacent dimensions, we introduce the Dirac-Bianconi operator \cite{bianconi2021topological}
\begin{equation}
	\mathcal{D} = \begin{pmatrix}
	0 & \mathbf{B}_1 \\
	\mathbf{B}_1^{\top} & 0
	\end{pmatrix}.
\end{equation}
This operator maps between 0- and 1-simplices, and it can be straightforwardly extended to higher dimensions. It plays a central role in describing higher-order dynamics, as it couples cochains across adjacent dimensions.

In this work, we choose the orientation that allows for the homogeneous state, i.e., $\mathcal{D}\cdot\vec{1}=\vec{0}$, where $\vec{1}=(1,...,1)^\top$ is a homogeneous vector of all ones, and $\vec{0}$ is the vector of all zeros. Note that this is not possible for every simplicial complex. For example, in the case of networks, only for Eulerian ones, i.e., those whose nodes have all even degree (see \cite{Giambagli2022diffusion} for details) admit the homogeneous state. Hence, those are the kinds of networks that we are going to consider.

By applying the Dirac-Bianconi operator \cite{bianconi2021topological} to the topological spinor, we can project the $0-$cochains onto the $1-$simplices and the $1-$coachains onto the $0-$simplices \begin{equation}
    \vec{\psi}=\mathcal{D}\vec{w}=\begin{bmatrix}
        \boldsymbol{B}_1\vec{v} \\  \boldsymbol{B}_1^\top\vec{u} 
    \end{bmatrix}.
\end{equation} The new topological spinor $\vec{\psi}$ has a vector of $1-$cochain as a first component, which are the $0-$cochains projected onto the $1-$simplices, and a vector of $0-$cochains as a second component. \\

Note that, by taking the square of the Dirac-Bianconi operator, one obtains the Hogde Laplacian, which is the operator governing higher-order diffusion \cite{bianconi2021topological}. As we will not deal with diffusion in this work, we will not discuss any further the Hodge Laplacian, whose detailed description can be found in the following Refs. \cite{bianconi2021higher,Lim2020}.

\section{Dirac-Bianconi driven oscillators}\label{sec:dirac_osc}

Let us consider a network of $N_n$ nodes and $N_l$ links, whose $N_n\times N_n$ adjacency matrix is given by $A$, and let us define the $N_n\times N_l$ boundary operator $\boldsymbol{B}_1$ through the orientation that admits the homogeneous state. Let us consider the $0$-cochain $\vec{u}$, column vector of dimension $N_n$, which denotes the topological signals in the space of the nodes, i.e., the $0$-simplices, and the $1$-cochain $\vec{v}$, column vector of dimension $N_l$, denoting the topological signals in the space of the links, i.e., the $1$-simplices. Note that each cochain has dimension\footnote{In principle, each $0$-cochain $u_i$ could be a $d$-dimensional vector, meaning that $\vec{u}$, the vector encoding all the $0$-cochains of the system, would have dimension $dN_n$. For instance, one could deal with $2$-dimensional $0$-cochains and $1$-dimensional $1$-cochains, as in \cite{muolo2024three}.} $1$. Going back to our system, the state variables of the whole network are described by the $N_n + N_l$-dimensional topological spinor $\vec{w}=(\vec{u}^\top , \vec{v}^\top)^\top$. The dynamics of the topological spinor is given by the following equation \begin{equation}\label{eq:dirac_spinor}
    \dot{\vec{w}}=\vec{F}(\vec{w},\mathcal{D}\vec{w}),
\end{equation} where $\vec{F}$ is an odd function.
 In terms of the state variables on nodes and links, this can be rewritten as 
\begin{equation}\label{eq:dirac1}
   \begin{bmatrix}
        \dot{\vec{u}}\\\dot{\vec{v}}
    \end{bmatrix}=\begin{bmatrix}
        \vec{f}(\vec{u},\boldsymbol{B}_1 \vec{v})\\ \vec{g}(\vec{v},\boldsymbol{B}_1^\top \vec{u})
    \end{bmatrix},
\end{equation} where $\vec{F}=(\vec{f}^\top , \vec{g}^\top)^\top$, and $f_i$ (resp. $g_j$) acts element-wise on the $0$-cochains $u_i,(\boldsymbol{B}_1 \vec{v})_i$, for $i=1,...,N_n$ (resp. $1$-cochains $v_j,(\boldsymbol{B}_1^\top \vec{u})_j$, for $j=1,...,N_l$).

Looking at the above equations, one can observe that the dynamical variables, i.e., the $0$- and $1$-cochains, interact with the variables on other simplices solely due to the Dirac-Bianconi operator. Otherwise, one would have $N_n$ decoupled systems on the nodes and $N_l$ decoupled systems on the links. This scheme is depicted in Fig. 1 for $N_n=N_l=3$. Note that this is different from the setting considered, for instance, in \cite{Giambagli2022diffusion}, where the presence of a diffusion coupling through the Hodge Laplacian was ensuring that the variables on the nodes were coupled to each other and the variables on the links as well, while the Dirac operator was responsible for the inter-dimensional coupling. Here, without the Dirac-Bianconi operator, there would be no coupling at all. 

\begin{figure}\label{fig:dirac_osc}
    \centering
    \includegraphics[width=\textwidth]{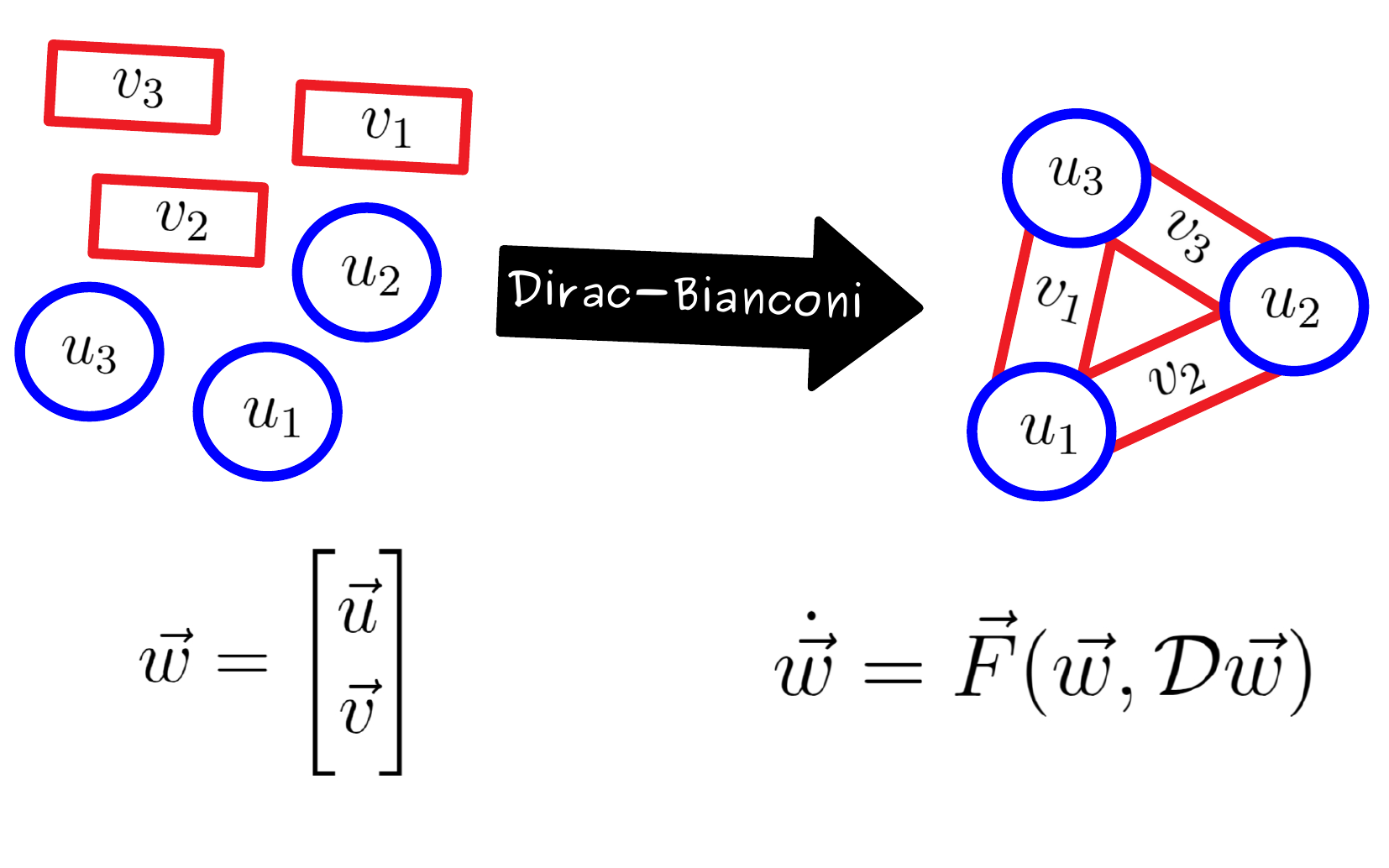}
    \caption{\textbf{Visualization of a Dirac-Bianconi driven oscillator:} the $0$-cochains $u_i$ and $1$-cochains $v_i$, grouped in the topological spinor $\vec{w}$, are isolated and do not feel each other's presence until the Dirac-Bianconi operator $\mathcal{D}$ couples them together, giving rise to a network whose dynamics is periodic. The oscillations are the results or the interactions between adjacent $0$- and $1$-cochains through the boundary and co-boundary operators $\boldsymbol{B}_1$ and $\boldsymbol{B}_1^\top$, i.e., the Dirac-Bianconi operator.}
\end{figure}

Let us now consider a solution of system \eqref{eq:dirac_spinor}, which we denote with $\vec{\mathcal{W}}$. We will call system \eqref{eq:dirac_spinor} a \textit{Dirac-Bianconi driven oscillator} when at least one of its solutions exhibit a stable periodic behavior, namely, when there exists a period $T$ such that $\vec{\mathcal{W}}(t+T)=\vec{\mathcal{W}}(t)$ for all time $t$. It is important to note that such periodic solution, which is driven by the Dirac-Bianconi coupling, is also a solution of the whole system. Each isolated unit, $0$- or $1$-cochain, does not exhibit a periodic behavior. In other words, we defined as a Dirac-Bianconi driven oscillator a system of decoupled inter-dimensional units which exhibit a stable periodic behavior when coupled via the Dirac-Bianconi operator. For the sake of simplicity, we have considered (and we will consider) the case of $0$- and $1$-cochains (i.e., nodes and links), but our definition can be extended to higher dimensional systems, i.e., $n$- and $n+1$-cochains. We chose the name Dirac-Bianconi driven oscillators to distinguish them from self-sustained oscillatory cochains, which are sometimes called \textit{topological oscillators}.  A topological oscillator is a self-sustained oscillator, meaning that it oscillates even without coupling, an example of such being the Topological Kuramoto \cite{millan2020explosive}, where each variable on the links (or any other simplex) is an oscillator. However, in our case, the oscillatory dynamics is induced by the Dirac-Bianconi operator, hence the name Dirac-Bianconi driven oscillator. \\
As anticipated in the Introduction, it was already observed that a pure Dirac-Bianconi coupling can yield patterns \cite{Giambagli2022diffusion}, which can also be time-periodic \cite{muolo2024three}. Indeed, the Dirac-induced patterns described in the latter Reference are an example of a Dirac-Bianconi driven oscillator, as it falls within the definition hereby given\footnote{Ref. \cite{muolo2024three} falls within the framework of Turing patterns \cite{Turing}, where a homogeneous equilibrium is needed to develop the theory. When the setting is extended to networks and higher-order structures \cite{muolo2024turing}, such conditions translates into considering identical oscillators and, if the state variables are cochains, additional conditions on the simplicial complex' structure and the basis in which the boundary operators are defined (see Ref. \cite{carletti2023global} for a detailed discussion).}. { There, a $2$-dimensional system lays in the nodes (i.e., two $0$-cochains on the $0$-chains) in a stationary state and oscillations were induced by the presence of third species on the links (i.e., a $1$-cochain on the $1$-chains). Note that, when more than one $d$-cochains lie on their respective $d$-chains, oscillations are possible for some given vector fields. In principle, one could further extend the framework of Ref.~\cite{muolo2024three} by considering a $2$-dimensional system on the nodes and another $2$-dimensional system on the links, both in a stationary state, and induce oscillations via the Dirac-Bianconi coupling. However, although the Dirac-Bianconi coupling is what triggers the emergence of oscillations, periodic behaviors would still be possible for appropriate values of the parameters. In this work, instead, we consider a setting in which any oscillatory dynamics would be impossible, if not for the Dirac-Bianconi coupling.}

\subsection{Example of a Dirac-Bianconi driven oscillator}

As an example, let us consider the following Dirac-Bianconi system on a $1$-dimensional simplicial complex, i.e., a network, of $N_n$ $0$-simplices, i.e., the nodes, and $N_l$ $1$-simplices, i.e., the links. We denote the $0$-cochains with the variables $u_i$, for $i=1,..,N_n$, and the $1$-cochains with $v_j$, for $j=1,...,N_l$. The equations are given by the following 
\begin{equation}
    \begin{cases}
        \label{eq:dirac_sys}
    \dot{u}_i=f_i(u_i,(\boldsymbol{B}_1\vec{v})_i),  \\ 
    \dot{v}_j=g_j(v_j,(\boldsymbol{B}_1^\top\vec{u})_j),
    \end{cases}
\end{equation}
 where all the $f_{i},g_{j}$ are differentiable functions, and at least one among them is nonlinear.

More specifically, the model we choose is inspired by the celebrated FitzHugh-Nagumo {(FHN)} model~ \cite{fitzhugh,nagumo}, a paradigmatic model in the study of neuronal dynamics~\cite{Murray2001}. Such model consists of two variables: the first is a fast variable (in the definition of slow-fast dynamics \cite{kuehn2015multiple}) describing the neuron's action potential through a cubic function, while the second is a slow variable modeling the ion density through a linear function\footnote{{Note that whether the FHN behaves as a fast-slow system depends on the time scale separation parameter, which we will denote with $\delta$ in what follows.}}. { Fast-slow systems are widely studied in mathematical neuroscience because the time scales of the firing of the neuron and the changes in the ion channels are very different. Moreover, such models exhibit rich dynamics, such as \textit{canard} cycles and Mixed Modes Oscillations (MMOs), which reproduce well the spiking of neurons~\cite{desroches2012mixed,desroches2016canards}. The {FHN} model can also exhibit periodic oscillations for certain parameter values and can be analyzed through phase reduction, both as a single oscillator and as a network \cite{nakao16,nakao2018phase,zhu2020phase}.} In our setting, we consider the fast variables as $0$-cochains, i.e., state variables on the nodes, while the slow variables as $1$-cochains, i.e., on the links. {Note that, while considering the $0$-cochains as the voltage of the neuron, the $1$-cochains do not find a direct correspondence with the recovery variable of the ion density of the node-based FHN model, for instance, because they are not segregated in the links. 
Indeed, this is not yet a realistic setting, given that, among other things, axons and dendrites are not symmetric links, but have a directionality. Nonetheless, we can think of $1$-cochains as a gradient slow variable, which triggers the spiking of the action potential; moreover, we believe that this can the starting point for modeling neuronal dynamics through this framework. The dynamical equations of the Dirac-Bianconi version of the FHN} model, which, for the sake of brevity, we will call DBFHN from now on, are given by the following {equations:}

\begin{equation}
    \begin{cases}
        \label{eq:DBFHN}
    \dot{u}_i=u_i-u_i^3-(\boldsymbol{B}_1\vec{v})_i+I_i,  \\ 
    \dot{v}_j=\delta_j((\boldsymbol{B}_1^\top\vec{u})_j-b_jv_j+\alpha_j a_j),
    \end{cases}
\end{equation} where $I_i,\delta_j,b_j,a_j$ are the parameters that depend on the respective local dynamics (on the node or on the link), and $\alpha_j=\alpha(\boldsymbol{B}_1^{(j)})=\pm 1$ is a coefficient that accounts for the change of orientation of the $j-$th link which appears in the $j-$th column of the boundary operator $\boldsymbol{B}_1^{(j)}$, so that $f_j(-v_j,(\boldsymbol{B}_1^\top\vec{u})_j)=-f_j(v_j,(\boldsymbol{B}_1^\top\vec{u})_j)$, meaning that the system is invariant with respect to the orientation of the links.

\begin{figure}[h!]\label{fig:DBFHN}
    \centering
    \includegraphics[width=\textwidth]{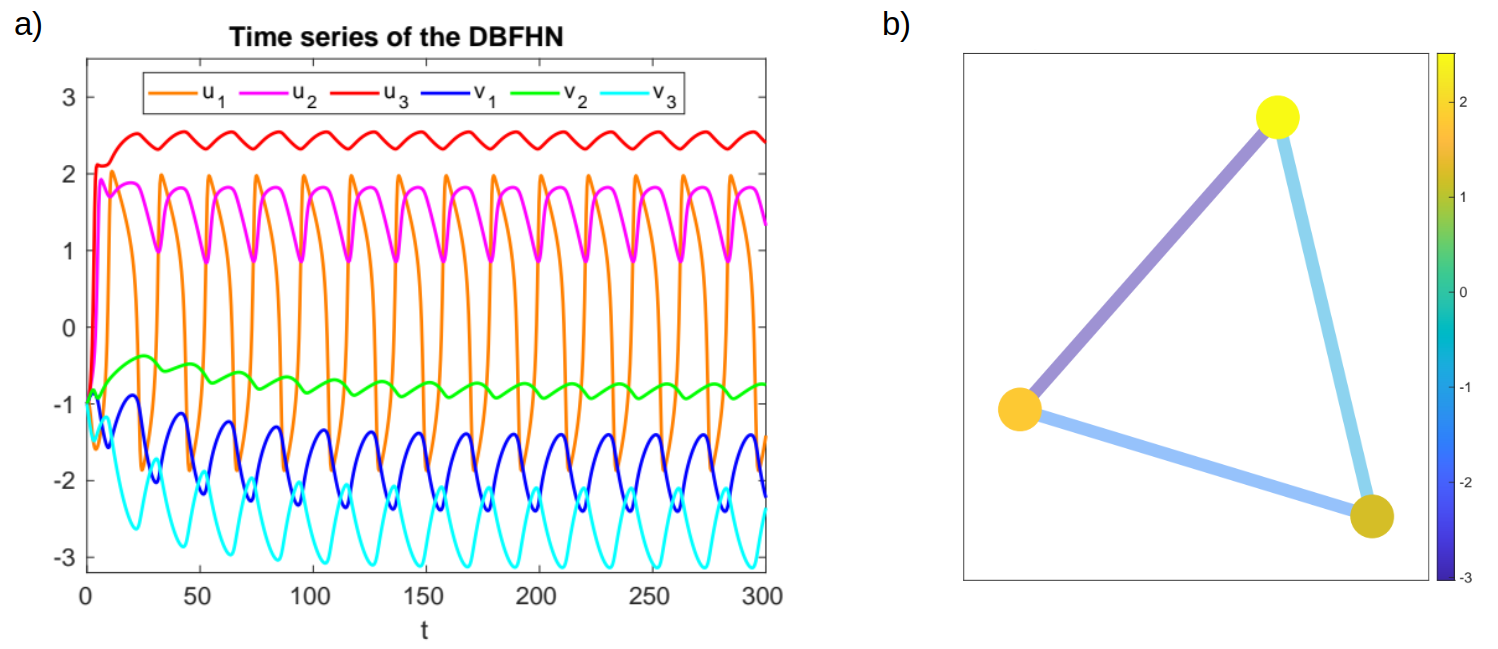}
    \caption{\textbf{Dynamics of the Dirac-Bianconi driven oscillator described in Eq. \eqref{eq:DBFHN}:} in panel a), we depict the time series of the $0$- and $1$-cochains; after a transient, the system follows a periodic trajectory of period $T=21.14$ time units (t.u.). Panel b) shows a snapshot of the values of the signals on nodes and links for $t=290$ t.u. The parameters are the following: $a_1=a_2=0.7$, $a_3=-3$, $b_1=b_2=b_3=0.3$, $\delta_1=\delta_2=\delta_3=0.08$, $I_1=I_2=I_3=0.8$. Note that $b_j$, $\delta_j$ and $I_i$ do not have to be equal for every system to have a periodic solution.}
\end{figure}

In Fig. 2, we show the time series of the DBFHN on a 1-simplicial complex (i.e., a network) of $N_n=3$ and $N_l=3$. We observe that, after a transient, the system settles onto a limit-cycle (i.e., a stable periodic trajectory), whose shape and period depend on the chosen parameters and the network structure. 

\begin{figure}[h!]\label{fig:period}
    \centering
    \includegraphics[width=\textwidth]{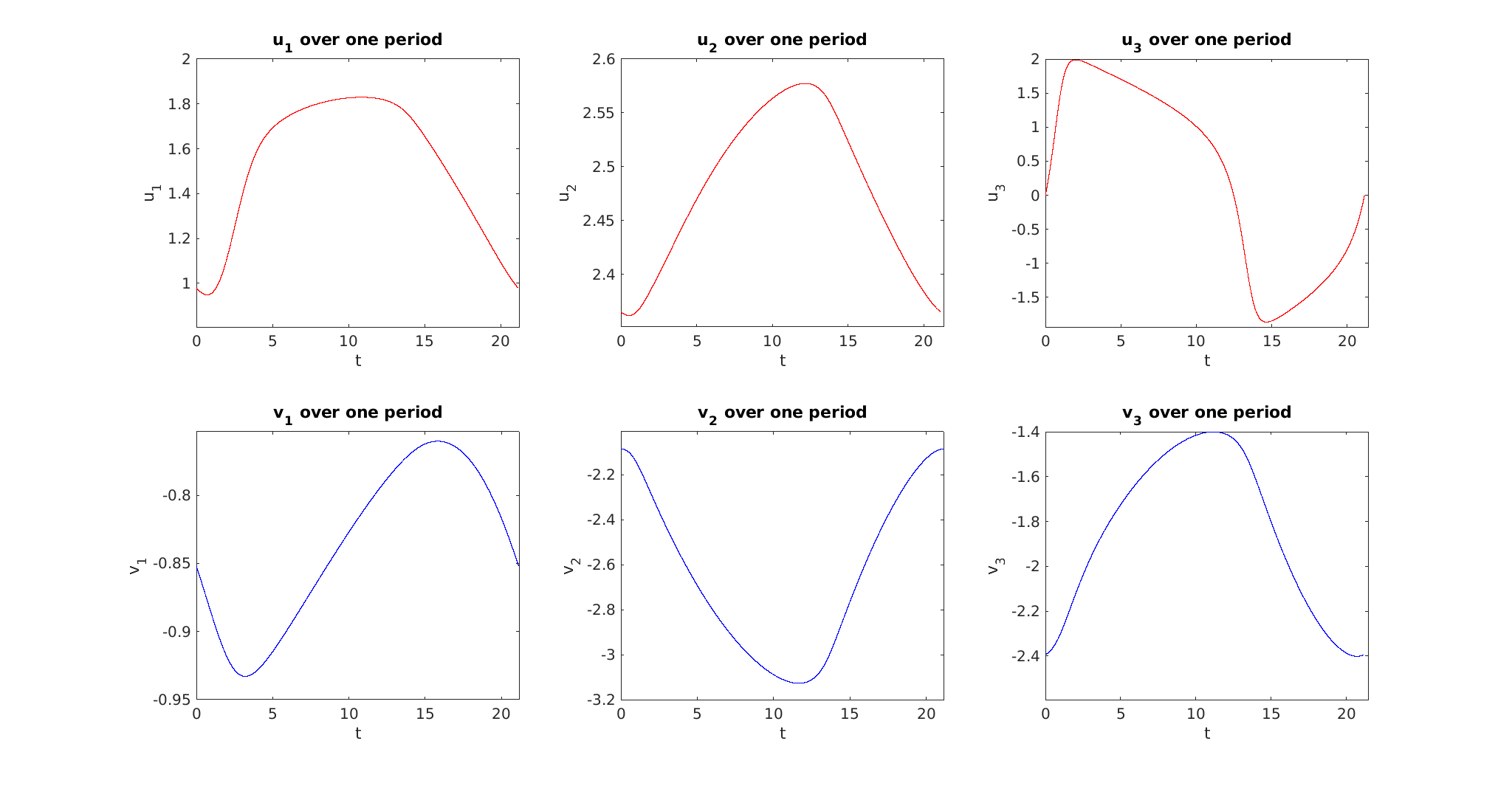}
    \caption{\textbf{Dynamics of each unit of the Dirac-Bianconi driven oscillator over one period:} $\boldsymbol{B}$ is such that the system has a homogeneous state and the model parameters are $a_1=a_3=0.7$, $a_2=-3$, $b_1=b_2=b_3=0.3$, $\delta_1=\delta_2=\delta_3=0.08$, $I_1=I_2=I_3=0.8$ (same setting of the previous Figure). The resulting period of the Dirac-Bianconi driven oscillator is $T=21.14$ t.u.}
\end{figure}

In Fig. 3, we show the time series of each variable of the DBFHN of Fig. 2 over one period. Let us remark that, although each variable has a periodic behavior, the limit cycle lies in the $N_n+N_l$ phase space. In fact, any cochain would not oscillate without the (Dirac-Bianconi) interaction with the others, and it is the whole system that behaves as an oscillator and not any single $0$- and $1$-cochain. Note that this approach, which has been considered on networks where oscillations arise only due to the interactions between the units \cite{wilson1972excitatory,brunel1999fast,kawamura2011collective,pazo2016quasiperiodic,devalle2017firing,nakao2018phase,politi2018collective,di2018transition,nakao2021sparse}, is radically different from what generally found in the literature on synchronization, where each unit (i.e., the system on the node) is an oscillator by itself and one studies how the interactions between many oscillators affect the global dynamics \cite{pikovsky2001synchronization,arenas2008synchronization}. 

As previously stated, this is a very simple setting: in fact, we have a $1$-simplicial complex of small size, the boundary operator with the orientation admitting a homogeneous state, only two state variables (one on the nodes and one on the links) and only one nonlinearity (for the dynamics of the $0$-cochain). One could extend this set up in several ways, for example, by considering nonlinear functions for every $s$-cochain, which could also be highly dimensional, more than $2$ kinds of cochains (e.g., on nodes, links and triangles), a more complex structure of the simplicial complex, or even cell-complexes, which are obtained with regular politopes instead of simplices~\cite{carletti2023global}, etc. Nonetheless, the substance of what we will present in the following Sections would not change. The simplicity of this setting in which collective oscillations are observed is aimed at highlighting the framework and the approach, which are the main contributions of this work.

\subsection{Coupled Dirac-Bianconi driven oscillators}

Let us now consider two DBFHNs, which we call DB1 and DB2, on two identical networks of $N_n=3$ and $N_l=3$, whose equations are, respectively,
\begin{equation}\label{eq:DBFHNs}
    \begin{cases}   
    \dot{\bar{u}}_i=\bar{u}_i-\bar{u}_i^3-(\bar{\boldsymbol{B}}_1\vec{\bar{v}})_i+\bar{I}_i,  \\ 
    \dot{\bar{v}}_j=\bar{\delta}_j((\bar{\boldsymbol{B}}_1^\top\vec{\bar{u}})_j-\bar{b}_j\bar{v}_j+\bar{\alpha}_j\bar{a}_j),
    \end{cases}~~~~~~~\mbox{ and }~~~~~~~ 
     \begin{cases}   
    \dot{\tilde{u}}_i=\tilde{u}_i-\tilde{u}_i^3-(\tilde{\boldsymbol{B}}_1\vec{\tilde{v}})_i+\tilde{I}_i,  \\ 
    \dot{\tilde{v}}_j=\tilde{\delta}_j((\tilde{\boldsymbol{B}}_1^\top\vec{\tilde{u}})_j-\tilde{b}_j\tilde{v}_j+\tilde{\alpha}_j\tilde{a}_j).
    \end{cases}
\end{equation} 

The boundary operators $\bar{\boldsymbol{B}}_1$ and $\tilde{\boldsymbol{B}}_1$ are identical and such that the orientation admits the homogeneous state, namely, \begin{equation}
    \bar{\boldsymbol{B}}_1=\tilde{\boldsymbol{B}}_1=\begin{bmatrix}
        1 & 0 &-1 \\ -1 & 1 & 0 \\ 0 & -1 & 1
    \end{bmatrix}.
\end{equation} Both systems have the same parameters as in the previous section with the only difference that $\tilde{I}_3=0.6$. With such parameters and without coupling, the period of DB1 is {$T_1$}$=21.140$ time units (t.u.), while that for DB2 is {$T_2$}$=20.415$ t.u., meaning that their oscillation frequencies are {$\omega_1$}$=0.2972$ t.u.$^{-1}$ and {$\omega_2$}$=0.3078$ t.u.$^{-1}$, respectively (hence, with a difference of $0.0106$ t.u.$^{-1}$). 

\begin{figure}\label{fig:coupled}
    \centering
    \includegraphics[width=\textwidth]{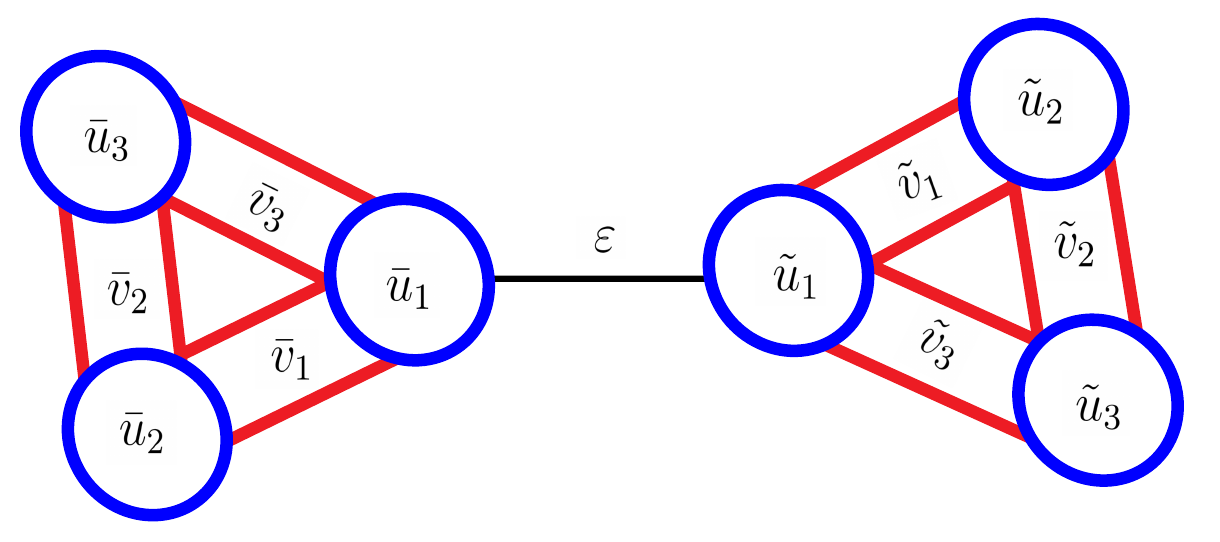}
    \caption{\textbf{Weak interaction between two Dirac-Bianconi driven oscillators:} pictorial representation of the setting considered in the text, namely, two Dirac-Bianconi driven oscillators weakly coupled via a pair of nodes. The pairwise coupling is, moreover, considered as linear and diffusive-like. Of course, more complex coupling configurations can be considered. Although it may seem that no higher-order effects take place in the coupling, synchronization between the two oscillators can be achieved only when considering the effects of the $1$-cochains, i.e., the variables on the links.}
\end{figure}

The two Dirac-Bianconi driven oscillators are connected via the nodes, as schematically described in Fig. 4. The equations for the two coupled DBFHNs are the following

\begin{equation} \label{eq:DBFHN1_incomplete}
    \begin{cases}   
    \displaystyle \dot{\bar{u}}_i=\bar{u}_i-\bar{u}_i^3-(\bar{\boldsymbol{B}}_1\vec{\bar{v}})_i+\bar{I}_i+\epsilon \bar{g}_i(\bar{u}_i,\tilde{u}_{i_1},...,\tilde{u}_{i_{N_n}})  \\ 
    \dot{\bar{v}}_j=\bar{\delta}_j((\bar{\boldsymbol{B}}_1^\top\vec{\bar{u}})_j-\bar{b}_j\bar{v}_j+\bar{\alpha}_j\bar{a}_j)
    \end{cases},
    \end{equation} 
    \begin{equation}\label{eq:DBFHN2_incomplete}
     \begin{cases}   
    \displaystyle \dot{\tilde{u}}_i=\tilde{u}_i-\tilde{u}_i^3-(\tilde{\boldsymbol{B}}_1\vec{\tilde{v}})_i+\tilde{I}_i+\epsilon \tilde{g}_i(\tilde{u}_i,\bar{u}_{i_1},...,\bar{u}_{i_{N_n}})  \\ 
    \dot{\tilde{v}}_j=\tilde{\delta}_j((\tilde{\boldsymbol{B}}_1^\top\vec{\tilde{u}})_j-\tilde{b}_j\tilde{v}_j+\tilde{\alpha}_j\tilde{a}_j)
    \end{cases},
\end{equation} where $\bar{g}_i,\tilde{g}_i$ are the coupling functions, and $\epsilon$ is the coupling strength, We have chosen the coupling through nodes as the simplest and, possibly, physically feasible way to couple Dirac-Bianconi driven oscillators. In fact, we consider each Dirac-Bianconi driven oscillator as an independent dynamical unit exhibiting complex (oscillatory) dynamics induced by the Dirac-Bianconi coupling, while the coupling between the oscillators is modeled with a classic diffusive coupling via the variables on the nodes ($0$-cochains). In principle, we could consider more elaborated higher-order couplings, however a pairwise weak ($~10^{-2}$) coupling allows us to focus on the dynamics of the Dirac-Bianconi driven oscillators. We assume the functions $\bar{g}_i$ and $\tilde{g}_i$ to be linear and diffusive. For the setting depicted in Fig. 4, where the coupled variables are $\bar{u}_1$ and $\tilde{u}_1$, the only non-zero coupling functions are $\bar{g}_i=\tilde{u}_1-\bar{u}_1=-\tilde{g}_i$. Of course, different coupling configurations could be considered, as well as different kinds of pairwise couplings, including nonlinear and non-diffusive ones. Again, we choose a simple setting in order to illustrate the phenomenon and the method. \\

As discussed previously, the two DBFHN oscillators have slightly different frequencies, meaning that an incoherent behavior is obtained without coupling, i.e., $\epsilon=0$. If we now set $\epsilon=10^{-2}$ and we consider Eqs. \eqref{eq:DBFHN1_incomplete} and \eqref{eq:DBFHN2_incomplete} coupled via only one node (as depicted in Fig. 4), we achieve synchronization only for large values of the coupling strength, as shown in Fig. 5, where we observe that the two oscillators synchronize their frequencies as the coupling becomes stronger. An analogous behavior is observed for other linear diffusive coupling combinations of different pairs of nodes (results not shown). {As depicted in Fig. 5, the system undergoes a saddle-node bifurcation~\cite{crawford1991introduction,strogatz2018nonlinear} with the coupling strength $\epsilon$ as the bifurcation parameter. In fact, when $\epsilon$ is large, there are at least two equilibrium points in the reduced phase model: one stable (a node), i.e., the synchronous state, and one unstable (a saddle). As the coupling strength diminishes, the two equilibrium points collide and annihilate each other. A more precise description of the unstable equilibrium point can be given via a phase description, as we will do in the next Section.}

\begin{figure}\label{fig:diff_coupl}
    \centering
    \includegraphics[scale=0.5]{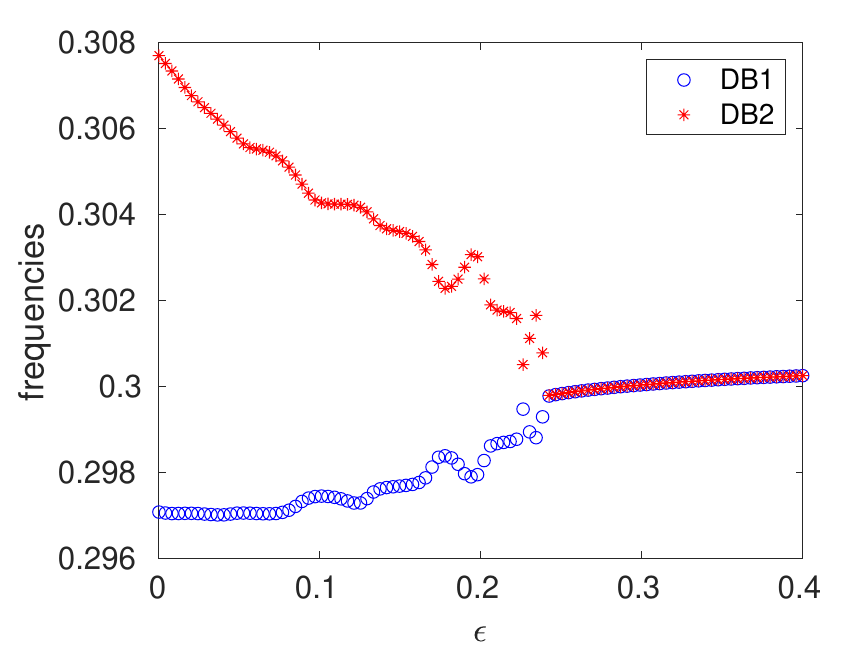}
    \caption{\textbf{Synchronization of two Dirac-Bianconi driven oscillators 
    {with weak diffusive coupling}:} we show the frequencies of DB1 (blue) and DB2 (red), coupled as in Eqs. \eqref{eq:DBFHN1_incomplete} and \eqref{eq:DBFHN2_incomplete} integrated until $t=6000$ t.u. and averaged over $100$ periods. We can observe that, with diffusive coupling, synchronization is achieved only for large values of the coupling strength. The parameters are $\bar{a}_1=\bar{a}_3=\tilde{a}_1=\tilde{a}_3=0.7$, $\bar{a}_2=\tilde{a}_2=-3$, $\bar{b}_1=\bar{b}_2=\bar{b}_3=\tilde{b}_1=\tilde{b}_2=\tilde{b}_3=0.3$, $\bar{\delta}_1=\bar{\delta}_2=\bar{\delta}_3=\tilde{\delta}_1=\tilde{\delta}_2=\tilde{\delta}_3=0.08$, $\bar{I}_1=\bar{I}_2=\bar{I}_3=\tilde{I}_1=\tilde{I}_2=0.8$ and $\tilde{I}_3=0.4$.}
\end{figure}

In addition to the diffusive coupling, let us now consider also a Dirac-Bianconi coupling. We can impose that the diffusive-like coupling between $0$-cochains has an effect on the $1$-cochains lying on the links (i.e., $1$-chains) adjacent to the coupled nodes (i.e., $0$-chains); vice versa, the $1$-cochains have an effect on the $0$-cochains lying on the nodes. If we look again at Fig. 4, this means that, via a Dirac-Bianconi coupling, $\bar{u}_1$ affects $\tilde{v}_1$ and $\tilde{v}_3$, and $\tilde{u}_1$ affects $\bar{v}_1$ and $\bar{v}_3$; vice versa $\tilde{v}_1$ and $\tilde{v}_3$ affect $\bar{u}_1$, and $\bar{v}_1$ and $\bar{v}_3$ affect $\tilde{u}_1$. We represent the higher-order coupling functions with $h^{\mathcal{D}}$.

\begin{equation}\label{eq:DBFHN1_complete}
    \begin{cases}   
    \displaystyle \dot{\bar{u}}_i=\bar{u}_i-\bar{u}_i^3-(\bar{\boldsymbol{B}}_1\vec{\bar{v}})_i+\bar{I}_i+\epsilon \bar{g}_i(\bar{u}_i,\tilde{u}_{i_1},...,\tilde{u}_{i_{N_n}})+\epsilon\bar{h}^{\mathcal{D}}_i \Big(\bar{u}_i,(\bar{\boldsymbol{B}}_1\vec{\tilde{v}})_{i_1},...,(\bar{\boldsymbol{B}}_1\vec{\tilde{v}})_{i_{N_n}} \Big),  \\ \\
   \displaystyle \dot{\bar{v}}_j=\bar{\delta}_j((\bar{\boldsymbol{B}}_1^\top\vec{\bar{u}})_j-\bar{b}_j\bar{v}_j+\bar{\alpha}_j\bar{a}_j)+\epsilon \bar{h}^{\mathcal{D}}_j\Big(\bar{v}_j,(\bar{\boldsymbol{B}}_1^\top\vec{\tilde{u}})_{j_1},...,(\bar{\boldsymbol{B}}_1^\top\vec{\tilde{u}})_{j_{N_l}}\Big),
    \end{cases}
    \end{equation} 
    \begin{equation}\label{eq:DBFHN2_complete}
     \begin{cases}   
    \displaystyle \dot{\tilde{u}}_i=\tilde{u}_i-\tilde{u}_i^3-(\tilde{\boldsymbol{B}}_1\vec{\tilde{v}})_i+\tilde{I}_i+\epsilon \tilde{g}_i(\tilde{u}_i,\bar{u}_{i_1},...,\bar{u}_{i_{N_n}})+\epsilon\tilde{h}^{\mathcal{D}}_i\Big(\tilde{u}_i,(\tilde{\boldsymbol{B}}_1\vec{\bar{v}})_{i_1},...,(\tilde{\boldsymbol{B}}_1\vec{\bar{v}})_{i_{N_n}}\Big),  \\ \\
   \displaystyle \dot{\tilde{v}}_j=\tilde{\delta}_j((\tilde{\boldsymbol{B}}_1^\top\vec{\tilde{u}})_j-\tilde{b}_j\tilde{v}_j+\tilde{\alpha}_j\tilde{a}_j)+\epsilon \tilde{h}^{\mathcal{D}}_j\Big(\tilde{v}_j,(\tilde{\boldsymbol{B}}_1^\top\vec{\bar{u}})_{j_1},...(\tilde{\boldsymbol{B}}_1^\top\vec{\bar{u}})_{j_{N_l}}\Big).
    \end{cases}
\end{equation} 

In this case, the system achieves synchronization even for weak coupling strengths and, moreover, as the coupling strength grows, the oscillations are slowed down, as shown in Fig. 6. {Observe that, qualitatively, the bifurcation behavior is the same as that of Fig. 5}{, but for a weaker coupling strength}. Hence, in our setting the Dirac-Bianconi coupling\footnote{Let us observe that the Dirac-Bianconi coupling of the previous equation would also emerge in the phase reduced model (see next Section) if the second order phase reduction \cite{leon19} of Eqs. \eqref{eq:DBFHN1_incomplete} and \eqref{eq:DBFHN2_incomplete} is performed, but it would be of order $\epsilon^2$.} is fundamental in achieving synchronization for weak couplings. In the next section we will explain this behavior through the theory of phase reduction.

\begin{figure}\label{fig:dirac_coupl}
    \centering
    \includegraphics[scale=0.5]{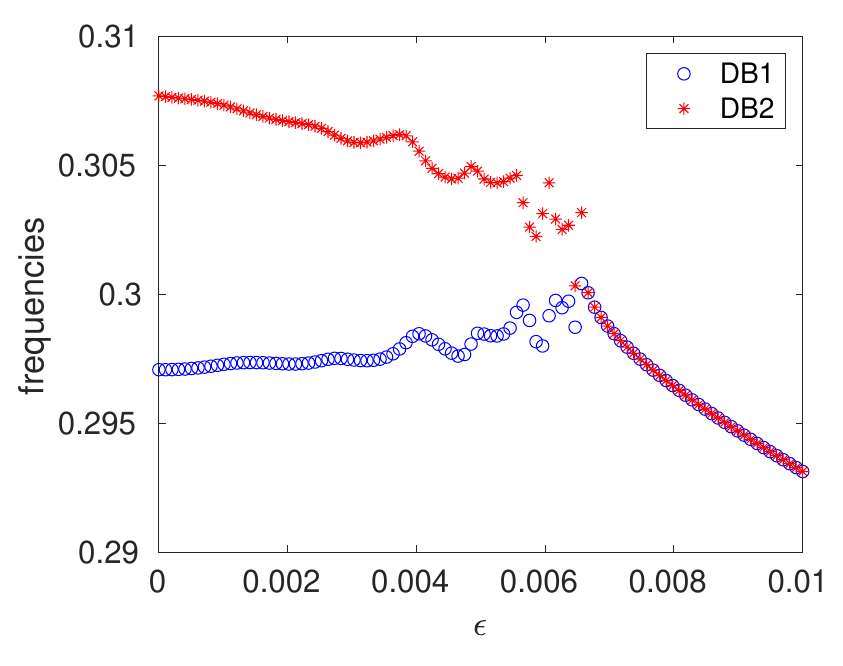}   
    \includegraphics[scale=0.5]{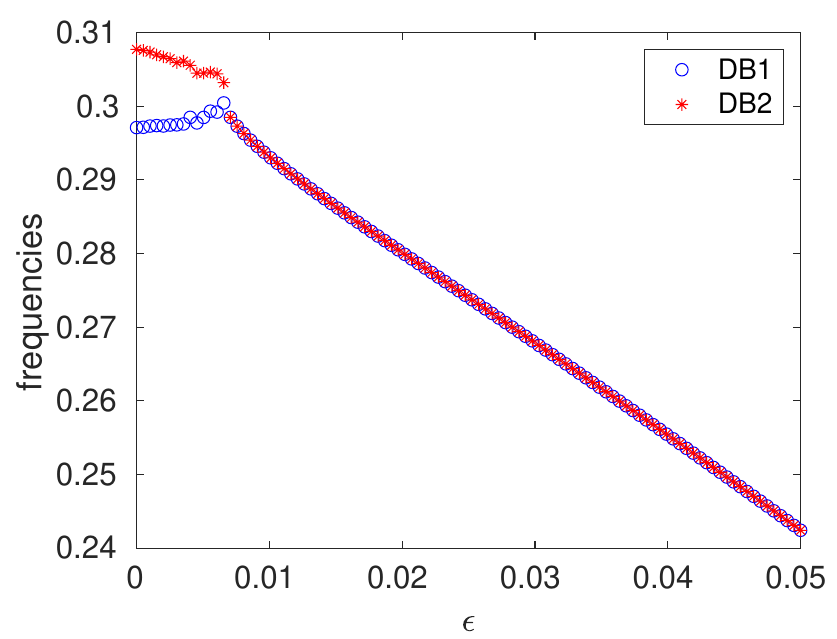}
    \caption{\textbf{Synchronization of two Dirac-Bianconi driven oscillators weakly coupled via the Dirac-Bianconi operator:} we show the frequencies of DB1 (blue) and DB2 (red), coupled as in Eqs. \eqref{eq:DBFHN1_complete} and \eqref{eq:DBFHN2_complete} integrated until $t=6000$ t.u. and averaged over $100$ periods. On the left panel, we can observe that synchronization is achieved for small values of the coupling. The right panel shows that the Dirac-Bianconi coupling slows down the oscillations as the coupling strength is increased. Note the different scales of the $x$ axis in the two panels. The parameters are $\bar{a}_1=\bar{a}_3=\tilde{a}_1=\tilde{a}_3=0.7$, $\bar{a}_2=\tilde{a}_2=-3$, $\bar{b}_1=\bar{b}_2=\bar{b}_3=\tilde{b}_1=\tilde{b}_2=\tilde{b}_3=0.3$, $\bar{\delta}_1=\bar{\delta}_2=\bar{\delta}_3=\tilde{\delta}_1=\tilde{\delta}_2=\tilde{\delta}_3=0.08$, $\bar{I}_1=\bar{I}_2=\bar{I}_3=\tilde{I}_1=\tilde{I}_2=0.8$ and $\tilde{I}_3=0.4$.}
\end{figure}

\section{Phase description of Dirac-Bianconi driven oscillators}\label{sec:phase_red_dirac}

We now briefly discuss classical phase reduction theory, which is a method to reduce the dynamics of an oscillator to a single variable, the phase $\vartheta$, using a perturbative expansion valid under weak coupling \cite{Kuramoto_book,nakao16,kuramoto2019concept,monga2019phase}. This method is widely applicable, since any self-sustained oscillator, i.e., with a stable limit cycle, can be reduced to a phase model if perturbations are small, and it has been recently extended to the case of (node-based) many-body interactions \cite{leon2025theory}. In this section we will first introduce the phase reduction and then focus on the phase sensitivity function, which allows us to interpret the numerical results of the previous section.

\subsection{Phase reduction approach}

Phase reduction is a perturbative dimensionality reduction technique useful in the study of coupled oscillators. It achieves a dimensionality reduction by describing the each oscillator in terms of a single variable, the phase $\vartheta$. The dynamics of a general population of $N$ oscillator is given by
\begin{equation}
    \dot{\vec{X}}_j = \vec{F}_j(\vec{X}_j) + \epsilon \sum_{k=1}^N\vec{g}_{jk}(\vec{X}_j,\vec{X}_k),
\end{equation}
where $\vec{F}_j$ is the velocity field of oscillator $j$, $\epsilon$ is the coupling strength and $\vec{g}_{jk}$ is the coupling function.

Because we are dealing with a population of oscillators, when uncoupled each subsystem displays an attractive limit cycle $\vec{X}_c(t)$. On the limit cycle we can define a phase variable $\vartheta$ with constant velocity $\dot{\vartheta} = \omega$. This definition can be extended to the whole basin of attraction with the concept of asymptotic phase and isochrons \cite{Win80}. This allows us to assign a value of the phase to every point in the basin of attraction, {$\vartheta(\vec{X}_j)$}. Applying the chain rule, we confirm its velocity is constant:
\begin{equation}
    \dot{\vartheta} = \nabla_{\vec{X}} \vartheta \cdot F(\vec{X}) = \omega.
\end{equation}

When the oscillators are coupled, the phase suffers variations in its velocity. Applying the chain rule:
\begin{equation}\label{eq:phase_model}
    \dot{\vartheta}_j = \omega_j + \epsilon \nabla_{\vec{X}_j} \vartheta(\vec{X}_j) \cdot \sum_{k=1}^N \vec{g}_{jk}(\vec{X}_j,\vec{X}_k).
\end{equation}

If the interactions are weak in comparison with the convergence to the limit cycle, i.e., $\epsilon\ll 1$, the oscillator always remains close to the limit cycle, allowing us to approximate $\vec{X} \approx \vec{X}_c$, yielding:
\begin{equation}\label{eq:phase_model_DB}
    \dot{\vartheta}_j = \omega_j + \epsilon \sum_{k=1}^N \vec{Z}_j(\vartheta_j) \cdot \vec{p}_{jk}(\vartheta_j,\vartheta_k) + \mathcal{O}(\epsilon^2),
\end{equation}
with $\vec{p}_{jk}(\vartheta_j,\vartheta_k)=\vec{g}_{jk}(\vec{X}_j(\vartheta_j),\vec{X}_k(\vartheta_k))$, and $Z_j(\vartheta_j) = \nabla_{\vec{X}} \vartheta(\vec{X}_c(\vartheta))$ the \textit{phase sensitivity function}, measuring phase response to infinitesimal perturbations \cite{winfree1967biological,Win80}. Stated differently, the phase sensitivity function measures the response of the oscillator to the effect of the interactions. Note that we have neglected $\mathcal{O}(\epsilon^2)$ terms~\footnote{When close to the bifurcation point, which is not the case considered in this work, such terms may have some relevant effects \cite{leon19,leon22a,leon20,bick2024higher}.}.

\subsection{Phase sensitivity function}

To numerically construct the phase model in Eq.~\eqref{eq:phase_model}, we need to determine the natural frequency $\omega$, the phase sensitivity function $Z(\vartheta)$, and the functional form of the limit cycle trajectory $\chi(\vartheta)$. These components are required to evaluate the interaction term $g_{jk}(\vec{X}_j, \vec{X}_k)$ and compute the coupling function $p_{jk}(\vartheta_j, \vartheta_k)$. However, analytical expressions for $Z(\vartheta)$ and $\chi(\vartheta)$ are typically available only for certain systems, such as the Stuart-Landau oscillator \cite{nakao16}, or the van der Pol oscillator \cite{Leon2023analytical}. 

In general, there are two main numerical approaches for computing the phase sensitivity function. The direct method, which involves applying small perturbations at different points along the limit cycle and measuring the resulting phase shifts. While conceptually straightforward, this method is computationally demanding, requiring careful control of the perturbation direction and magnitude to stay within the linear regime. A more elegant approach is given by the adjoint method, which consists in solving the adjoint equation \cite{ermentrout1996type,hoppensteadt2012weakly,brown2004phase}, with the advantage of avoiding repeated perturbations. The adjoint equation is given by:
    \begin{equation}
        \frac{dZ(\vartheta)}{dt} = -J(\vartheta)^\top Z(\vartheta),
    \end{equation}
    where $J(\vartheta)$ is the Jacobian matrix of the vector field $F(\vec{X})$ evaluated along the limit cycle $\chi(\vartheta)$. The solution $Z(\vartheta)$ must be $T$-periodic and satisfy the normalization condition:
    \begin{equation}
        Z(\vartheta) \cdot \frac{d\chi(\vartheta)}{dt} = \omega,
    \end{equation}
    ensuring that the phase grows uniformly with time along the limit cycle. The interested reader may consult \cite{nakao16,monga2019phase} for a tutorial on this method and its implementation.

\subsection{Phase sensitivity function of a Dirac-Bianconi driven oscillator}

\begin{figure}[h!]\label{fig:PSF}
    \centering
    \includegraphics[width=\textwidth]{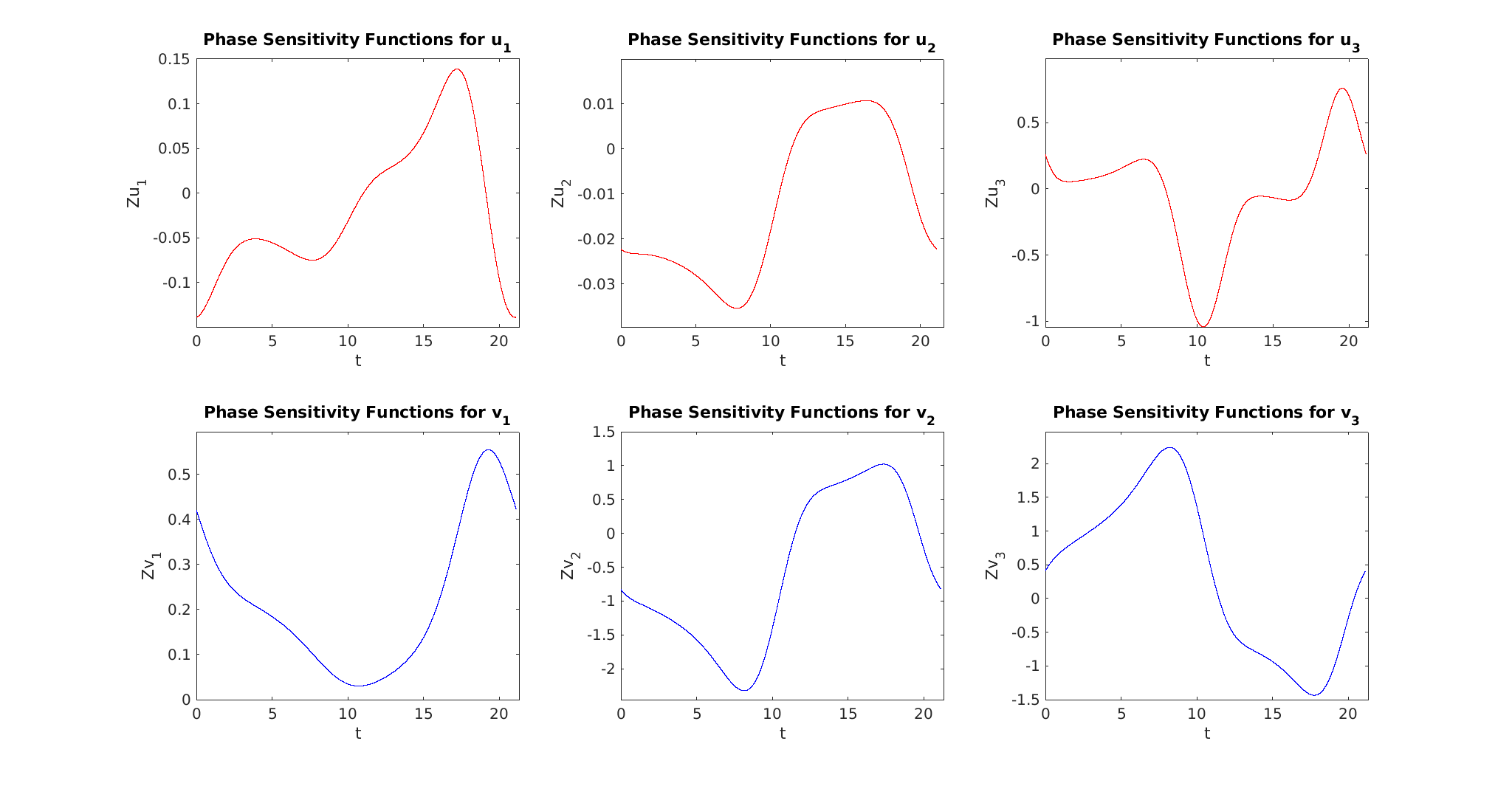}
    \caption{\textbf{Phase sensitivity functions of the Dirac-Bianconi driven oscillator described in Eq. \eqref{eq:DBFHN}:} we can see that the phase sensitivity functions of the variables on the nodes are small, while those of the variables on the links are larger. This explains why a Dirac-Bianconi coupling is necessary to achieve synchronization for weak coupling strengths. The parameters are the following: $a_1=a_3=0.7$, $a_2=-3$, $b_1=b_2=b_3=0.3$, $\delta_1=\delta_2=\delta_3=0.08$, $I_1=I_2=I_3=0.8$. Observe that the phase sensitivity function of $v_3$ is higher than the others, because $a_3$, which is the excitability in the FHN model, is higher. Due to the fact that $v_3$ interacts with $u_1$ and $u_3$, we see that also their phase sensitivity functions are higher than that of $u_2$, which is not directly interacting with $v_3$. With reference to Eq. \eqref{eq:DBFHNs} and to the Figs. of the previous section, note that the phase sensitivity functions of the two oscillators are similar, given that the two oscillators have similar parameters and frequencies.}
\end{figure}

To understand the numerical results shown in the previous section, namely, why the two coupled oscillators synchronize better through a Dirac-Bianconi coupling, we compute the phase sensitivity functions of the two oscillators using the adjoint method described above. This method allows us to quantify how sensitive the phase of the oscillator is to small perturbations in each of its variables. The model setup and parameter values are the same as in the numerical example, so the two oscillators have similar dynamics and nearly identical frequencies. In Fig. 7, we display the phase sensitivity functions for all variables of one Dirac-Bianconi driven oscillator. It is clear that the phase sensitivity functions associated with the node variables are relatively small. On the other hand, the phase sensitivity functions associated with the link variables are larger. This indicates that interactions with respect to the link variables have a much stronger influence on the timing of the oscillations. This observation provides an explanation for why the Dirac-Bianconi coupling is better to achieve synchronization in this system, particularly when the coupling strength is weak. Since the link variables have a much greater effect on the phase, a coupling that acts through them is more effective at aligning the phases of the oscillators, thus yielding synchronization. A coupling acting only on the node variables is, indeed, insufficient, because their influence on the phase is too small. This is clearly shown in Fig. 5, where the coupling needs to be large in order for the frequencies to synchronize. Finally, by computing the phase sensitivity functions for different Dirac-Bianconi systems, one can gain useful insights into which variables have a greater effect on the phase. This makes it possible to determine, in advance, the most suited coupling configuration to achieve synchronization. In this way, the phase sensitivity function serves as a practical tool for both guiding the design of effective coupling strategies and control the dynamics in systems of Dirac-Bianconi driven oscillators.

{ \subsection{Phase coupling functions}

Let us conclude by exploiting the phase reduction to further understand the synchronization dynamics of coupled Dirac-Bianconi driven oscillators and quantitatively explain the bifurcation causing the emergence of synchronization for strong enough couplings shown in Figs. 5 and 6. \\
After having performed the phase reduction, we obtain a phase equation for each of the two Dirac-Bianconi driven oscillators, DB1 and DB2, analogous to Eq.~\eqref{eq:phase_model_DB}: \begin{equation}
    \begin{cases}
    \dot{\vartheta}_1 = \omega_1 + \epsilon \vec{Z}_1(\vartheta_1) \cdot \vec{p}_{12}(\vartheta_1,\vartheta_2),\\
    \dot{\vartheta}_2 = \omega_2 + \epsilon \vec{Z}_2(\vartheta_2) \cdot \vec{p}_{21}(\vartheta_2,\vartheta_1).
    \end{cases}
\end{equation}

As explained in~\cite{nakao16}, under the hypothesis that $\omega_1 - \omega_2 \sim O(\epsilon)$, namely, $\omega_1=\omega+\Delta_1$ and $\omega_2=\omega+\Delta_2$, with $\Delta_{1, 2} \sim O(\epsilon)$, we can introduce a slow phase for each DB oscillator, namely, $\varphi_1=\vartheta_1-\omega t$ and $\varphi_2=\vartheta_2-\omega t$. If we then describe the above system in terms of the slow phases, we obtain, after averaging approximation, 
\begin{equation}
    \begin{cases}
    \dot{\varphi}_1 = \Delta_1 + \epsilon \Gamma_{12}(\varphi_1-\varphi_2),\\
    \dot{\varphi}_2 = \Delta_2 + \epsilon \Gamma_{21}(\varphi_2-\varphi_1),
    \end{cases}
\end{equation} where \begin{equation}
    \Gamma_{ij}=\frac{1}{T_i}\int_{0}^{T_i}\vec{Z}_i(\varphi_i+\omega \tau)\cdot \vec{p}_{ij}(\varphi_i+\omega \tau,\varphi_j+\omega \tau)d\tau.
\end{equation}

Finally, defining $\phi=\varphi_1-\varphi_2$, and $\Delta_1-\Delta_2=\Delta\omega$, we obtain the equation of the phase difference \begin{equation}\label{eq.phdif}
    \dot{\phi}=\Delta\omega+\epsilon\Big(\Gamma_{12}(\phi)-\Gamma_{21}(-\phi)\Big)=\Delta\omega+\epsilon\Gamma(\phi).
\end{equation}

\begin{figure}\label{fig:dirac_coupl}
    \centering
    \includegraphics[scale=0.5]{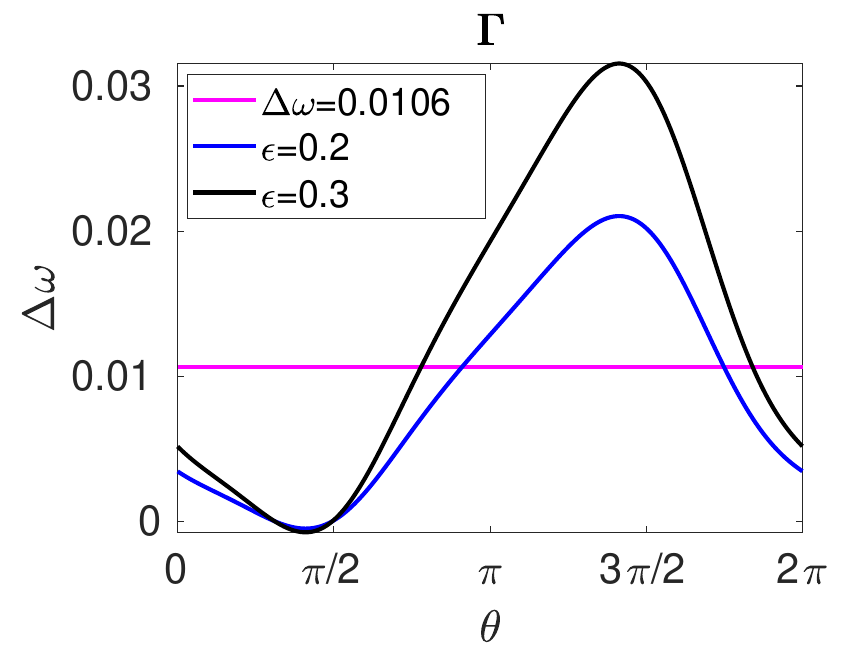}   
    \includegraphics[scale=0.5]{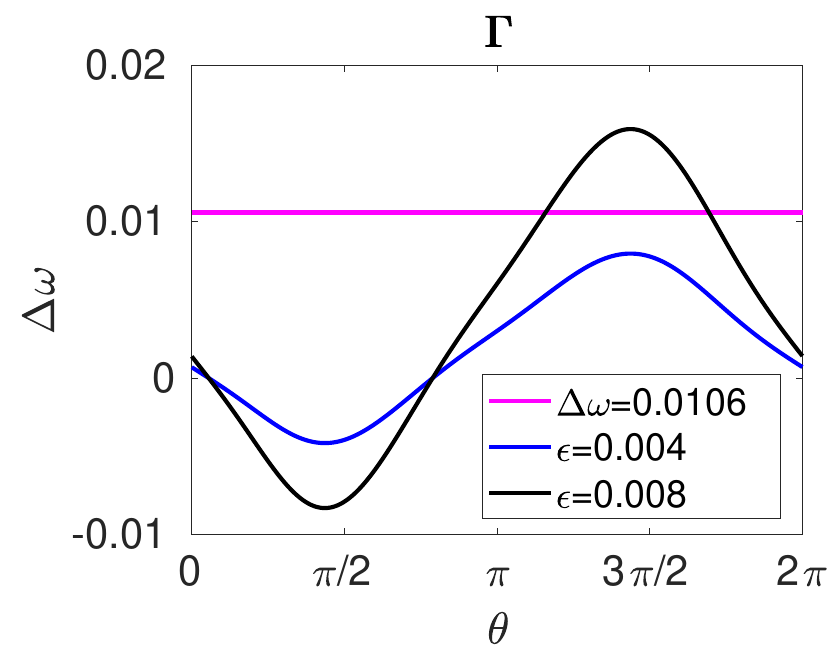}
    \caption{{\textbf{Dynamics of the reduced phase models of two coupled Dirac-Bianconi driven oscillators:} we show the dynamics of the relative phase of the two Dirac-Bianconi driven oscillators corresponding to the settings of Fig. 5 (left panel) and Fig. 6 (right panel). The magenta curves represent the frequency difference of the two uncoupled oscillators; the blue curves indicate the $\Gamma$ function for couplings such that synchronization is not achieved, i.e., no fixed point in the dynamics of the relative phases; the black curves represent the $\Gamma$ function for couplings such that synchronization is achieved, i.e., two fixed points, given by the intersection of the black and magenta curves, one stable (corresponding to the synchronous phase lock state, one unstable). We observe that, while the phase description accurately describes the case with Dirac-Bianconi coupling (right panel, with reference to Fig. 6), it only provides approximate results in the case with diffusing coupling (right panel, with reference to Fig. 5). Indeed, the phase reduction predicts synchronization for $\epsilon$ well below $0.2$, while direct numerical simulations of the system (Fig. 5) show that synchronization emerges above that coupling strength. The reason for this discrepancy is that the phase reduction is valid as long as the coupling is weak. This is why the description of the case with Dirac-Bianconi coupling is accurate (synchronization is obtained for $\varepsilon \gtrsim 0.006$), while the case with diffusive coupling not (synchronization is obtained for $\varepsilon \gtrsim 0.2$). The parameters are $\bar{a}_1=\bar{a}_3=\tilde{a}_1=\tilde{a}_3=0.7$, $\bar{a}_2=\tilde{a}_2=-3$, $\bar{b}_1=\bar{b}_2=\bar{b}_3=\tilde{b}_1=\tilde{b}_2=\tilde{b}_3=0.3$, $\bar{\delta}_1=\bar{\delta}_2=\bar{\delta}_3=\tilde{\delta}_1=\tilde{\delta}_2=\tilde{\delta}_3=0.08$, $\bar{I}_1=\bar{I}_2=\bar{I}_3=\tilde{I}_1=\tilde{I}_2=0.8$ and $\tilde{I}_3=0.4$.}}
\end{figure}

The function $\Gamma$ allows us to describe the dynamics of the phase difference of the two oscillators, as shown in Fig. 8. In the figure, we depict the $\Gamma$ function for the case of diffusive coupling (left panel) and Dirac-Bianconi coupling (right panel). We observe that, when the coupling strength is high enough, $\Gamma$ intersects $\Delta\omega$ giving rise to two fixed points, see Eq.\eqref{eq.phdif}, one stable and one unstable. However, comparing Fig. 8 with Figs. 5 and 6, we see that, while the case of Dirac-Bianconi coupling of Fig. 6 is well described by the phase approximation of Fig. 8 (right panel), this approximation fails in describing the case where the coupling is diffusive (see Fig. 8, left panel and Fig. 5). In fact, from the phase description it would appear that a coupling strength of $\epsilon=0.15$ would be enough for the system to synchronize, while this is not the case. This discrepancy occurs because the coupling strength necessary to synchronize in the case with diffusive coupling is not small, hence, we are not in the range of validity of the phase reduction at first order. In fact, with higher values of the coupling strength, second-order effects are non-negligible and more accurate results can be obtained with a second-order phase reduction, as done in~\cite{leon19,Mau23,bick2024higher}. On the other hand, the Dirac-Bianconi coupling enables synchronization for much smaller values of the coupling parameters, allowing us to obtain an accurate description by means of the phase reduction.
}

\section{Conclusion}\label{sec:the_end}

In this work we have introduced and described the dynamics of Bianconi-Dirac driven oscillators. These are higher-order networks in which the variables, called topological signals (or cochains), lie on the nodes, links, triangles, and higher-order simplicies and are coupled via the Dirac-Bianconi operator, which drives the system into an oscillatory behavior. 
We have fully described a simple Dirac-Bianconi driven oscillator, whose dynamics is inspired by the well-known FitzHugh-Nagumo neural model, in which the structure is solely made of nodes and links. {We have put the fast variable on the nodes, while the slow variable on the links, and we have shown that, thanks to the Dirac-Bianconi coupling, such structure can exhibit a periodic dynamics.} Then, by coupling two slightly different oscillators, we have shown that the classic diffusive coupling needs to be strong in order for the oscillators to synchronize, while, if the coupling is also mediated by the Dirac-Bianconi operator, synchronization is observed also for weak coupling. To understand this behavior, we derived a full phase description of a Dirac-Bianconi driven oscillator through phase reduction theory and obtained the phase sensitivity function, which measures the response of the phase to the interactions on nodes and links. We have shown that, for our example, the slow variables on the links have higher phase sensitivity functions than the fast variables on the nodes, meaning that a Dirac-Bianconi coupling is more fitting to achieve synchronization for weak values of the coupling. In future works, { more realistic models could be considered. For example, one could use this approach to model the dynamics of macro-region of the brains: each region could be represented by the nodes and their voltage by $0$-cochains, while the links with their $1$-cochains would stand for the flow of current. One may also consider $2$-simplices (i.e., triangles) to model the dynamics of $2$-cochains, which would represent the magnetic field (a bivector~\cite{dorst2007geometric}) induced by the flow of current.  Moreover,} the concept of Dirac-Bianconi driven oscillator could be extended to directed simplicial complexes \cite{gong2024higher} and higher-order multiplex networks \cite{krishnagopal2023topology}, while different dynamics could be considered, such as stochastic systems \cite{zhu2022phase,zhu2025complex} or slow-fast systems whose phase sensitivity function could be approximated analytically \cite{Leon2023analytical}. To achieve synchronization, optimization techniques could be applied with respect to the coupling connectivity \cite{nakao2021sparse}, input function \cite{kato2021optimization}, or coupling functions \cite{namura2024optimal}, as well as control methods \cite{d2023controlling,coraggio2026controlling}. Our method, that we have made explicit for a simple case, can be straightforwardly extended to an arbitrary number of coupled Dirac-Bianconi driven oscillators of any dimensions and can provide a useful modeling tool in brain dynamics, where edge-signals are becoming increasingly relevant \cite{faskowitz2022edges}.

\section*{Acknowledgements}
R.M. acknowledges JSPS KAKENHI 24KF0211 for financial support. I.L. acknowledges support by Grant No.~PID2021-125543NB-I00, funded by MICIU/AEI/10.13039/501100011033 and by ERDF/EU. Y.K. acknowledges JSPS KAKENHI JP22K14274 and JST PRESTO JPMJPR24K3 for financial support. H.N. acknowledges JSPS KAKENHI 25H01468, 25K03081, and 22H00516 for financial support.
R.M. is grateful to Future University Hakodate for the hospitality. The authors are grateful to Ginestra Bianconi for discussions and for the invitation to contribute to this special issue.

\section*{Authors contribution}
Conceptualization: R.M., Y.K.; Methodology: R.M., I.L., H.N.; Formal Analysis and Investigation: R.M.; Visualization: R.M.; Numerical Implementation: R.M.; Writing—original draft preparation: R.M.; Writing—review and editing: all authors; Project Coordination: R.M.; Funding acquisition: Y.K., H.N.

\section*{References}

\end{document}